\documentclass[prl,reprint,superscriptaddress]{revtex4-1}

\usepackage{graphicx,hyperref}
\usepackage{amsmath}
\usepackage{amssymb}
\usepackage[titletoc]{appendix}
 
\usepackage{bm}
\usepackage{amsfonts}
\usepackage{epstopdf}
\usepackage{xcolor}
\usepackage{color}
\usepackage{wasysym}

\usepackage{subcaption}
\hypersetup{
	colorlinks,    citecolor=blue,    filecolor=blue,    linkcolor=blue,    urlcolor=blue
}

\usepackage{verbatim}

\date{\today}

\begin{document}

\begin{abstract}

We present the first 3D particle-in-cell simulations of laser driven sheath-based ion acceleration in a kilotesla-level applied magnetic field. 
The applied magnetic field creates two distinct stages in the acceleration process associated with the time-evolving magnetization of the hot electron sheath and results in a focusing, magnetic field-directed ion source of multiple species with strongly enhanced energy and number.
The benefits of adding the magnetic field are downplayed in 2D simulations, which strongly suggests the feasibility of observing magnetic field effects under experimentally relevant conditions.

\end{abstract}

\title{Generation of focusing ion beams by magnetized electron sheath acceleration}

\author{K. Weichman}
\affiliation{Department of Mechanical and Aerospace Engineering, University of California at San Diego, La Jolla, CA 92093, USA}
\author{J.J. Santos}
\affiliation{University of Bordeaux, CNRS, CEA, CELIA, UMR 5107, F-33405 Talence, France}
\author{S. Fujioka}
\affiliation{Institute of Laser Engineering, Osaka University, Osaka 565-0871, Japan}
\author{T. Toncian}
\affiliation{Institute for Radiation Physics, Helmholtz-Zentrum Dresden-Rossendorf e.V., 01328 Dresden, Germany}
\author{A.V. Arefiev}
\affiliation{Department of Mechanical and Aerospace Engineering, University of California at San Diego, La Jolla, CA 92093, USA}

\vskip -2.0cm
\maketitle
\vskip -3.0cm

Recent advances in all-optical magnetic field generation have made experimentally accessible new regimes of magnetized high energy density physics (HEDP) relevant to applications including 
inertial fusion energy \cite{strozzi2012fast_ignition,fujioka2016fast_ignition,sakata2018isochoric}
and laboratory astrophysics \cite{bulanov2015labastro,albertazzi2014jet}.
In particular, the introduction of laser-driven coil targets \cite{fujioka2013coil,santos2015coil,santos2018coil,gao2016coil,goyon2017coil} capable of generating nanosecond-duration, hundreds of Tesla to kilotelsa-level magnetic fields over 100's of microns
at currently-existing large laser facilities including ILE \cite{fujioka2013coil}, LULI \cite{santos2015coil,santos2018coil}, and OMEGA \cite{gao2016coil,goyon2017coil}
introduces new possibilities in magnetized, relativistic laser-produced plasma.
The understanding of the impact of strong magnetic fields on HEDP is rapidly evolving and has spurred research in areas including electron beam transport \cite{johzaki2015transport,bailly2018guiding},
laser-produced magnetic reconnection \cite{fiskel2014reconnection}, 
and ion acceleration \cite{arefiev2016protons,kuri2018rpa_cp,cheng2019rpa}.

In particular the ion acceleration induced by the expansion of a laser-heated electron sheath into vacuum \cite{gurevich1966expansion,mora2003expansion} 
presents an attractive platform for the study of magnetic field effects in laser-produced plasmas.
Following its initial demonstration \cite{clark2000tnsa,maksimchuk2000tnsa,snavely2000tnsa},
non-magnetized sheath-based ion acceleration has been extensively studied \cite{macchi2013tnsa_review},
including in configurations compatible with experimental magnetic field generation platforms.
Improvements in the ion source characteristics are additionally desirable for applications including isochoric heating \cite{patel2003isochoric} 
and ion fast ignition \cite{roth2001ion_fast}.
It is therefore advantageous to elucidate the mechanism via which applied magnetic fields can beneficially alter sheath-based ion acceleration, particularly in the context of realistic magnetic field strengths.

Given the computational expense associated with 3D simulations, it would seem desirable to study the effect of the applied magnetic field in the context of 2D simulations. 
The limitations of using 2D simulations to represent 3D physics are well known in the target normal sheath acceleration (TNSA \cite{wilks2001tnsa}) regime (i.e. without an applied magnetic field), with, for example, the conclusion that 2D simulations over-predict both the acceleration time and the maximum ion energy (e.g. Refs. \cite{sgattoni2012_2D-3D,dhumieres2013_2D_3D}).
However, the addition of the magnetic field as a new element in sheath-based ion acceleration requires re-evaluating the appropriateness of 2D simulations (such as those presented in Refs.~\onlinecite{arefiev2016protons,santos2018coil}) to study what is inherently a 3D phenomenon.

In this Letter, we present the first 3D simulations of sheath-based ion acceleration with a kilotesla-level applied magnetic field. 
We demonstrate that the magnetization of hot electrons results in a two-stage ion acceleration process producing a focusing, magnetic field-directed ion source of multiple species with strongly enhanced energy and number. 
We show that electron magnetization is tied to the balance of thermal to magnetic pressure in the hot electron sheath (plasma $\beta_e$), which changes over the course of the acceleration and in turn drives a fundamental change in the sheath electric field configuration.
We additionally find that the beneficial effects of the applied magnetic field are substantially downplayed in 2D simulations, on which basis we predict the feasibility of observing the acceleration mechanism we describe under experimentally relevant conditions.

We simulate a relativistically intense laser pulse interacting with the preplasma in front of a solid density plastic (CH) target with and without an applied magnetic field in 2D and 3D using the particle-in-cell code EPOCH \cite{arber2015epoch}. The magnetic field strength and laser spot size were chosen to make 3D simulations tractable below machine-scale, which necessitated a 2000~T magnetic field. 
We also investigate the ability of 2D simulations to reproduce the magnetic field benefits observed in 3D at 2000~T. Following this analysis, we conduct additional 2D simulations with a 400~T field and larger laser spot to probe the relevance of the ion acceleration process we observe in 3D under experimentally realizable conditions.
These 2D simulations had similar computational cost as the 3D simulations. 
Unless explicitly stated, all simulation results were obtained from 3D simulations.

\begin{figure*}
    \includegraphics[width=0.95\linewidth]{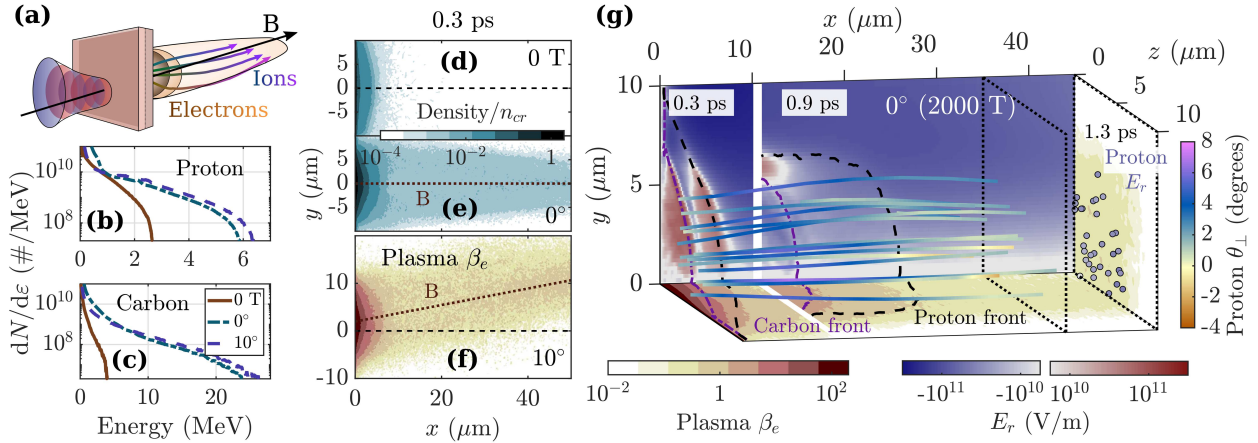}
    \caption{Ion acceleration with a strong applied magnetic field. (a)~Schematic of simulation setup. An intense laser pulse impacts the preplasma at the front surface of a solid density target, heating electrons and accelerating ions from the back surface. (b)-(c)~Final energy spectra for (b)~protons and (c)~carbon ions. (d,e)~Electron density for (d)~TNSA (no magnetic field) and (e)~2000~T target normal magnetic field. (f)~Plasma $\beta_e$ calculated from the density with 2000~T magnetic field directed at $10^\circ$ in the $x-y$ plane. (g)~Proton trajectories, ion front locations, radial electric field, and plasma $\beta_e$ at different times for target normal 2000~T magnetic field. The protons shown have final energy above 4~MeV.}
    \label{fig:combined}
\end{figure*}

The simulation setup is shown schematically in Fig. \ref{fig:combined}a. 
The laser pulse has a wavelength of 1.06~$\mu$m, is spatially and temporally Gaussian with a 150~fs FWHM duration and a 3~$\mu$m FWHM spot size (both given in terms of the intensity), and has a peak intensity of $2\times10^{19}$~W/cm$^2$. The preplasma has an exponential profile with a scale length ($1/e$ density falloff) of 1.5~$\mu$m. We model the solid density target as a 5~$\mu$m thick slab of fully ionized proton and carbon plasma with number density $n_p = n_C = 10 n_{cr}$, where $n_{cr} \equiv m_e \omega_0^2/4\pi e^2$ is the critical density associated with the reflection of a laser pulse with frequency $\omega_0$. The laser is linearly polarized in the $y$-plane, propagates in the $x$-direction, and is focused onto the front target surface. We apply a static uniform magnetic field of $B_0 = 2000$~T at either target normal incidence (0$^\circ$, $B_x = B_0$) or angled upward at 10$^\circ$ in the $x$-$y$ plane. 
The simulation domain is $90\times37\times24$~$\mu$m for the largest simulation, which we resolve with 30 cells/$\mu$m in $x$ and 20 cells/$\mu$m in $y$ and $z$. 
Electrons, protons, and carbon ions are represented by 10, 5, and 5 cubic B-spline macroparticles per cell through most of the domain, with 20 macroparticles per cell for protons and carbon ions within 0.5~$\mu$m of the target rear surface. We set $x=0$ to correspond to the target rear surface. For convenience, $t=0$ denotes the time when the peak of the laser pulse would impact the front target surface. The simulation is run until the highest energy protons begin to leave the simulation box. 

Sheath-based ion acceleration is driven by hot electrons.
When the laser interacts with the front surface preplasma, it generates a population of hot electrons which stream through the target and establish a sheath field on the rear surface, which then accelerates ions.
We find that the applied magnetic field does not substantially alter the laser-produced electron energy or angular spectrum. In our cases, the applied magnetic field is weak compared to the peak laser magnetic field and the electron gyro-frequency is low compared to the laser frequency, which precludes the resonant heating effect observed at substantially higher magnetic field strength \cite{arefiev2015electrons,kuri2018rpa_cp,cheng2019rpa}. 

Although there is no apparent difference in the laser-produced hot electrons, we find that the accelerated ion energy and number, especially for the heavier ion species, are substantially enhanced by the application of the 2000~T field (Figs. \ref{fig:combined}b and \ref{fig:combined}c).
This enhancement can be traced to the magnetic field restricting the transverse spread of hot electrons within the solid density target (akin to Ref. \onlinecite{bailly2018guiding}). 
For our simulation parameters, the laser spot size is comparable to the hot electron Larmor radius $\rho_L = cp_\perp/eB$, where we estimate $c p_\perp \sim T$ by the slope temperature $T \approx 0.8$~MeV ($\mbox{e}^{-\varepsilon/T}$ fit).
The magnetic field reduces the hot electron transport across field lines and increases the sheath electron density (e.g. Fig. \ref{fig:combined}d versus Fig. \ref{fig:combined}e) and accelerating electric field. 

We additionally find that the magnetic field fundamentally changes the electric field configuration of the sheath \textit{through the magnetization of hot electrons}, resulting in high energy ions which are 
1)~magnetic field-directed (the angular spectrum peaks along the field direction), and 2)~magnetic field-focusing (coming to a focus along the field line).
Qualitatively, electrons are magnetized when the magnetic force dominates the electric force perpendicular to the field lines, i.e.
\begin{equation}
    \left|e\mathbf{v}\times \mathbf{B}\right|/c > \left|e\mathbf{E_\perp}\right|,
\end{equation}
which requires at a minimum $B>E_\perp$.
We estimate the sheath electric field generated by hot electrons as
\begin{equation}
    E_\perp \approx 4 \pi |e| n_e \lambda_{De} = \sqrt{4 \pi n_e T},
\end{equation}
where $\lambda_{De} \equiv \sqrt{T/4\pi e^2 n_e}$ is the hot electron Debye length corresponding to the local electron density $n_e$.
We estimate $B \approx B_0$ (the diamagnetic effect does not change the order of magnitude of $B$).
The comparison between the electric and magnetic fields in the sheath is approximately
\begin{equation}
    E_\perp/B \sim \rho_L/\lambda_{De} \sim \sqrt{\beta_e},
\end{equation}
where $\beta_e \equiv 8 \pi n_e T/B^2$ is the ratio of thermal to magnetic pressure. 
Thus we monitor $\beta_e$, which is associated with the collective processes of sheath formation and magnetic pressure, to infer the electron magnetization in the sheath.

\begin{figure}
    \centering
    \includegraphics[width=0.95\linewidth]{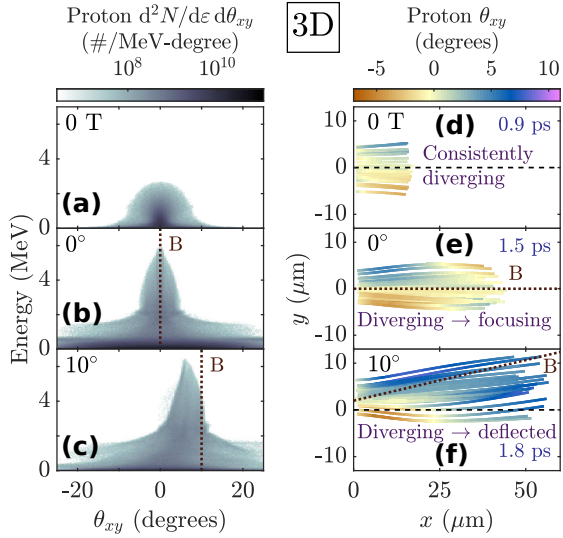}
    \caption{Ion focusing and deflection with applied magnetic field. (a)-(c)~Proton angular energy distribution. (d)-(f)~Angle between proton forward ($x$) momentum and $y$-momentum, projected on the $x-y$ plane. (a),(d)~TNSA (no magnetic field). (b),(e)~Target normal 2000~T magnetic field. (c),(f)~2000~T magnetic field directed at $10^\circ$ in the $x-y$ plane. The protons shown have final energy above 2~MeV (0~T case) or 4~MeV (2000~T cases).}
    \label{fig:six}
\end{figure}

Close to the target surface and during the initial stage of acceleration, ions see an electron population with $\beta_e \gg 1$. During this stage, the ions quickly gain the majority of their final energy, but there is no difference in the electric field configuration relative to a typical unmagnetized sheath.

\begin{figure}
    \centering
    \includegraphics[width=0.95\linewidth]{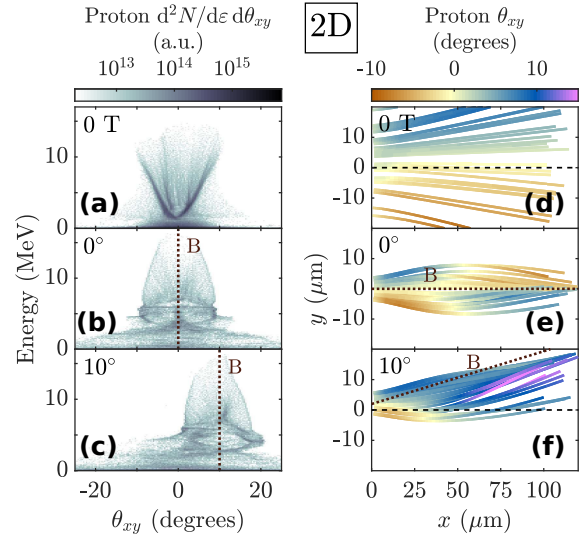}
    \caption{Ion focusing and deflection in 2D simulations. (a)-(c) Proton angular energy distribution. (d)-(f) Angle between proton forward ($x$) momentum and $y$-momentum, projected on the $x-y$ plane. (a),(d)~TNSA (no magnetic field). (b),(e) Target normal 2000~T magnetic field. (c),(f) 2000~T magnetic field directed at $10^\circ$ in the $x-y$ plane. The protons shown have final energy above 10~MeV.}
    \label{fig:2D_six}
\end{figure}

Farther from the target surface and during the second stage of acceleration, high energy protons and carbon ions encounter an electron population with $\beta_e < 1$. This population is magnetic field-following (i.e. magnetized), which we demonstrate directly by tilting the magnetic field by $10^\circ$ in the $x$-$y$ plane, e.g. in Fig. \ref{fig:combined}f. 
We find that the net deflection of electrons from the target-normal direction in the $10^\circ$ case causes the high energy ion population to be deflected as well (Figs. \ref{fig:six}c and \ref{fig:six}f).
While the protons are still in the process of deflecting toward the field lines at the end of our 3D simulation, 2D simulations demonstrate that the high energy ion population becomes fully magnetic field-directed (Fig. \ref{fig:2D_six}c).

The magnetization of hot electrons additionally induces ion focusing about the magnetic field lines (e.g. Fig. \ref{fig:six}e).
The target-transverse, outward directed electric field experienced by ions during the initial, unmagnetized stage of the acceleration ($x<10$ $\mu$m in Fig. \ref{fig:combined}g) results from the relative mobility of electrons, which are able to expand past the ions in both the target-normal and target-transverse directions.
In contrast, in the magnetized sheath, the magnetic field energy exceeds the electron thermal energy ($\beta_e <1$) and the electrons become less mobile than the ions in the magnetic field-transverse direction, reversing the transverse sheath configuration.
We find that as ions pass $\beta_e \lesssim 0.5$, the electric field becomes magnetic field-focusing
(e.g. $x>10$ $\mu$m in Fig. \ref{fig:combined}g).
This focusing effect persists over a long time and visibly pulls high energy ions toward the field lines (e.g. Figs. \ref{fig:combined}g and \ref{fig:six}e,f). It also reduces the angular spread of high energy ions relative to the 0~T case (Fig. \ref{fig:six}a-c).
This sheath field reversal-induced focusing is in contrast with the shock-mediated collimation proposed to explain the formation of astrophysical jets (e.g. Ref.~\onlinecite{albertazzi2014jet}).

Although we observe the magnetic field-deflecting and magnetic field-focusing effects of electron magnetization in 2D simulations (Fig. \ref{fig:2D_six}), we find that 3D simulations are required to accurately capture the benefits of adding the magnetic field.
This may be due in part to fundamental differences in physical processes which affect the strength of the sheath electric field in 2D versus 3D geometry.
First, in 3D the hot electron sheath expands in two transverse directions, while in 2D it only expands in one, meaning the accelerating electric field drops less in 2D than in 3D for expansion over the same distance.
Second, in 3D the electrostatic potential well created by charge separation has finite depth and allows sufficiently hot electrons to carry kinetic energy out of the system, while in 2D, the electrostatic potential does not converge as the hot electrons move away from the ions, meaning even very hot electrons can transfer their full kinetic energy into sheath potential energy.
The effect of these differences can be seen even in TNSA (no magnetic field), where it is well-known that 2D simulations over-predict the ion energy (see, for instance the difference in peak energy in Figs. \ref{fig:six}a and \ref{fig:2D_six}a). 

\begin{figure}
    \centering
    \includegraphics[width=0.95\linewidth]{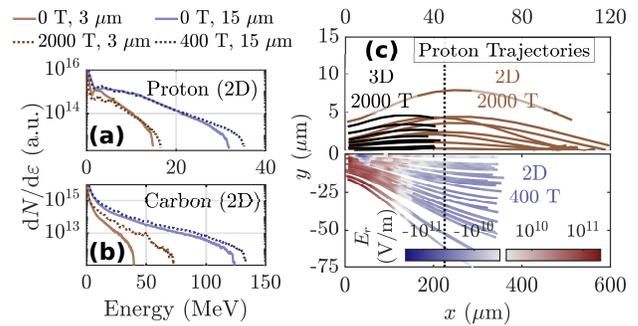}
    \caption{Comparison of ion acceleration for 2000~T cases with 3~$\mu$m focal spot (2D and 3D) and 400~T case with 15~$\mu$m focal spot (2D only). (a) Proton and (b) carbon ion energy spectra from 2D simulations. (c, top) 2000~T: comparison of proton trajectories in 2D and 3D. (c, bottom) 400~T: radial electric field seen by protons plotted along trajectories. Dotted line denotes (top)~$x=45$~$\mu$m, roughly where $p_y$ changes sign in the 2000~T 2D case, and (bottom)~$x=5\times45=225$~$\mu$m. The protons shown have final energy above 4~MeV (3D case), 10~MeV (2D, 2000~T), or 25~MeV (2D, 400~T).
    }
    \label{fig:2D_comp}
\end{figure}

The addition of a sufficiently strong magnetic field modifies both the transverse expansion of the hot electron sheath and the behavior of the (now magnetized) electrons escaping the potential well, and clearly degrades the fidelity of 2D simulations.
In a series of otherwise identical 2D simulations (same magnetic field strength, laser spot size, etc as the 3D cases), we observe that 2D simulations downplay the energy enhancing, deflecting, and focusing effects of the magnetic field.
2D simulations fail to reproduce the substantial energy enhancement observed in 3D simulations, e.g. the factor of 2 and 5 increases in the peak ion energy shown for proton and carbon ions in Figs. \ref{fig:combined}b,c, respectively, and instead predict almost no enhancement in the peak proton energy and only a moderate increase in the peak carbon energy (brown lines in Figs. \ref{fig:2D_comp}a,b). 
Additionally, we find that 2D simulations substantially over-predict the distances ions must propagate to be deflected towards the magnetic field lines and subsequently focused (e.g. Fig.~\ref{fig:2D_comp}c,top).

When 3D simulations are not tractable, for example at lower magnetic field strength, we can leverage the property that 2D simulations downplay the effects of the applied magnetic field to predict whether the magnetic field can still beneficially impact ion acceleration.
Fig. \ref{fig:2D_comp}c (red/blue trajectories) shows the development of a persistent focusing electric field for a 2D simulation where we have decreased the magnetic field strength and increased the laser spot size by a factor of 5 ($B_x = 400$~T, 15~$\mu$m FWHM; keeping the ratio between the spot size and the Larmor radius roughly fixed). 
The transition from radially outward to radially inward electric field associated with the electron magnetization occurs later and the focal length is longer in the 400~T case than the 2000~T case, even when the distances are scaled by a factor of 5 (as in the visual comparison between the top and bottom panels of Fig.~\ref{fig:2D_comp}), but the transition to focusing behavior is still clearly seen, on which basis we expect the benefits of adding the magnetic field to be observable in 3D at experimentally relevant field strengths. 

In summary, the net result of adding a strong magnetic field is a magnetic field-directed, magnetic field-focusing ion source of multiple species with enhanced energy and number. The ion acceleration process features a fundamental change in the sheath dynamics mediated by the electron magnetization and occurs in two stages, an initial target normal stage with high energy gain and high divergence driven by electrons which are unmagnetized in the sheath but transversely confined through magnetization in the target, followed by a subsequent stage of ion deflection and focusing in the magnetic field direction driven by magnetized electrons. 
We term this two stage ion acceleration process magnetized electron sheath acceleration (MESA).
We have additionally demonstrated that the benefits of adding the magnetic field are downplayed in 2D simulations, on which basis we predict the relevance of MESA under experimentally relevant conditions.

This research was supported in part by the DOE Office of Science under Grant No. DE-SC0018312. 
K.W. was supported in part by the DOE Computational Science Graduate Fellowship under Grant No. DE-FG02-97ER25308. 
J.J.S. acknowledges the support of the "Investments for the future" program IdEx Bordeaux LAPHIA (ANR-10-IDEX-03-02).
Particle-in-cell simulations were performed using EPOCH \cite{arber2015epoch},  developed under UK EPSRC Grant Nos. EP/G054940, EP/G055165, and EP/G056803.
This work used HPC resources of the Texas Advanced Computing Center (TACC) at the University of Texas at Austin and the National Energy Research Scientific Computing Center (NERSC), a U.S. Department of Energy Office of Science User Facility operated under Contract No. DE-AC02-05CH11231.
Data collaboration was supported by the SeedMe2 project \cite{chourasia2017seedme} (http://dibbs.seedme.org).


\bibliographystyle{ieeetr}

\end{document}